\documentclass[fleqn,twoside]{article}
\usepackage{espcrc2}
\input{psfig}

\usepackage{graphicx}
\usepackage[figuresright]{rotating}
\newcommand{\be}{\begin{equation}}
\newcommand{\ee}{\end{equation}}
\newcommand{\bea}{\begin{eqnarray}}
\newcommand{\eea}{\end{eqnarray}}
\newcommand{\nn}{\nonumber}

\newcommand{\AmS}{{\protect\the\textfont2
  A\kern-.1667em\lower.5ex\hbox{M}\kern-.125emS}}

\title{On three-body B decays to charm}
\author{P. Colangelo\address[INFN]{Istituto Nazionale di Fisica Nucleare,
Sezione di Bari, \\ 
via Amendola n.173, I-70126 Bari, Italy}
}
       
\begin{document}

\begin{abstract}
I briefly describe the use of the three-body decay
modes $B^0 \to D^{(*)-} D^{(*)0} K^+$ to investigate open issues 
in charmed meson spectroscopy, and of the time dependent 
$B^0 (\overline {B^0}) (t) \to D^{-} D^{+} \pi^0$ transitions 
for a measurement of $\cos (2 \beta)$.
\vspace{1pc}
\end{abstract}
\maketitle

\section{$B^0 \to D^{(*)-} D^{(*)0} K^+$ AND CHARMED MESON SPECTROSCOPY}

Important physical information can be obtained analyzing nonleptonic 
many-body $B$ decays, once clean channels have been recognized and selected. 
Remarkable examples are the three-body modes into pairs 
of charmed mesons and a light pseudoscalar meson, for which one can attempt a 
theoretical treatment based on heavy quark and chiral light meson symmetries
\cite{Colangelo:2002dg}.
A few of such modes have been recently observed. Here I consider  
\begin{eqnarray}
B^0 &\to& D^{*-} D^{(*)0} K^+    \label{channel1}\\
B^0 &\to& D^{-}  D^{(*)0} K^+    \label{channel2}
\end{eqnarray}
observed  by the BABAR \cite{Aubert:2001wz} and BELLE \cite{belle}
collaborations with a (preliminary) measurement of the branching fractions.
Such modes are useful for shedding light on open issues
of the charmed meson spectroscopy; from the theoretical point of view, 
they can be used to assess, e.g., the validity of the factorization ansatz.

\begin{figure}[htb]
\hspace*{0.3cm}
\psfig{figure=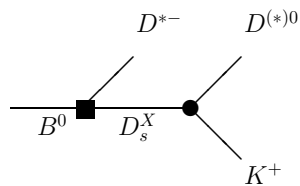,height=5.0 cm}
\vspace*{-1.0cm}
\caption{Polar diagram contributing to the modes 
$B^0 \to  D^{*-}  D^{(*)0} K^+$. The box represents a weak transition, the dot
a strong vertex.}
\label{fig:diagrams}
\end{figure}

The main idea  is that 
the decays (\ref{channel1}) and (\ref{channel2}) essentially
proceed through two-body intermediate states, 
$B^0 \to D^{(*)-} D_s^X$,  followed by the strong transition  
$D_s^X \to D^{(*)0} K^+$ (fig.\ref{fig:diagrams}),
the intermediate  $D_s^X$ being charmed strange mesons. 
%
%
%
In the factorization approach \cite{review_fact}, the $B^0 \to D^{(*)-} D_s^X$
amplitudes are expressed as the product of the semileptonic $B^0 \to D^{(*)-}$ 
matrix element and the $D_s^X$ current-vacuum matrix element.
When the infinite charm quark mass limit is exploited, the only contributions
with non-vanishing  $D_s^X$ current-vacuum matrix elements 
correspond to the states $D_s^X=D_s^*$ and $D_{s0}$ 
(with their radial excitations) for 
$B^0 \to D^{*-} D^0 K^+$, and  $D_s^X=D_s^*$, 
$D_s$  and $D^*_{s1}$ (together with their radial excitations)
for $B^0 \to D^{*-} D^{*0} K^+$. Analogous selection rules hold for
$B^0 \to D^{-} D^{(*)0} K^+$. $D_{s0}$ and $D^*_{s1}$ are charmed mesons
with  $J^P=0^+,1^+$, respectively,
belonging to the $s_\ell^P={1\over 2}^+$ heavy meson $(\bar s c)$ doublet,
$s_\ell^P$ being the spin-parity of the light degrees of freedom in the meson.
Such states are still unobserved, therefore it is worth identifying suitable
channels for individually studying their features. 

In order to calculate the amplitudes in fig.\ref{fig:diagrams}
one needs the  effective couplings parametrizing the vertices
$D^0 K^+D_s^*$, $D^0 K^+ D_{s0}$, $D^{*0} K^+ D_s^*$, $D^{*0} K^+ D_s$,
$D^{*0} K^+ D^{*+}_{s1}$ 
(and analogous vertices involving radial $D_s^X$ resonances).
In the heavy quark limit, all the couplings can be expressed in
terms of two independent coupling constants, $g$ and $h$ for negative and 
positive parity $D_s^X$ states, respectively.
This can be shown considering the effective QCD lagrangian, with
heavy quark spin-flavour and light quark chiral symmetries
\cite{hqet_chir}, that
describes the interactions of heavy negative and 
positive parity mesons with the light pseudoscalar mesons:
\bea
{\cal L}_I &=& i \;
g \; Tr\{H_b \gamma_\mu \gamma_5 {\cal A}^\mu_{ba} {\bar H}_a\} \nn \\
&+&  \;[ \; i \;
h \; Tr\{H_b \gamma_\mu \gamma_5 {\cal A}^\mu_{ba} {\bar S}_a\}
+ \; h.c. \;] \;\; .  \label{L}
\eea
The fields $H_a$ in (\ref{L})
describe the $J^P=(0^-,1^-)$ ${\bar q }Q$ mesons with
$s_\ell^P= {1\over 2}^-$:
\begin{equation}
H_a = \frac{(1+{\rlap{v}/})}{2}[P_{a\mu}^*\gamma^\mu-P_a\gamma_5] \;\;,
\label{neg}
\end{equation}
the operators $P^{*\mu}_a$ and $P_a$ respectively annihilating the 
$1^-$  and $0^-$  mesons of four-velocity $v$ ($a=u,d,s$ is a light flavour 
index). Analogously, the  fields $S_a$ describe positive parity
$s_\ell^P= {1\over 2}^+$  states:
\begin{equation}
S_a = \frac{(1+{\rlap{v}/})}{2}[P_{a\mu}^{\prime *}\gamma^\mu \gamma_5 - 
P^\prime_a] \;\;.
\label{pos}
\end{equation}
The octet of the light pseudoscalar mesons is included in (\ref{L}) through
the field $\displaystyle \xi=\exp\big({ i {\cal M} \over f_\pi}\big)$, with
$f_{\pi}=131 \; MeV$ and $\cal M$ given by
\begin{equation}
\left (\begin{array}{ccc}
\sqrt{\frac{1}{2}}\pi^0+\sqrt{\frac{1}{6}}\eta & \pi^+ & K^+  \\
\pi^- & -\sqrt{\frac{1}{2}}\pi^0+\sqrt{\frac{1}{6}}\eta & K^0 \\
K^- & {\bar K}^0 &-\sqrt{\frac{2}{3}}\eta
\end{array}\right ) . \nonumber \label{M}
\end{equation}
Moreover,
${\cal A}_{\mu ba}=\frac{1}{2}\left(\xi^\dagger\partial_\mu \xi-\xi
\partial_\mu \xi^\dagger\right)_{ba}$.
\par
It is straightforward to derive the relations of 
$g_{D^*_s D K}, g_{D^*_s D^* K}$ and $g_{D^* D_s K}$ to the effective 
coupling $g$,
and of $g_{D_{s0} D K}$ and $g_{D^*_{s1} D^* K}$ to $h$, using 
(\ref{L})-(\ref{M}). 
The  full widths of the intermediate states can also be expressed 
in terms of $g$ and $h$, in the approximation of dominance of two-body
decay modes. The masses of positive parity mesons
can be taken from theoretical 
determinations: $m_{D_{s0}}\simeq m_{D^*_{s1}}=m_{D_s}+\Delta$, with
$\Delta \simeq 0.5 \,GeV$  \cite{noidelta}. 

The calculation of $B \to D^{(*)} D_s^X$,
in the factorization approach, requires  the semileptonic 
$B^0 \to D^{(*)-}$ matrix element and the $D_s^X$  decays constant.
In the heavy quark limit, the former is given
in terms of the Isgur-Wise function $\xi_{IW}$,
while  the decay constants 
$f_{D_s}$ and $f_{D^*_s}$, as well as $f_{D_{s0}}$ and $f_{D^*_{s1}}$,
are simply related \cite{DeFazio:2000up}.
Other parameters, CKM elements and QCD coefficients,
appear in the same combination as in the two-body
$B^0 \to D^{*-} D_s^{*}$ amplitude, the branching ratio
of which is rather accurately  known \cite{Groom:in}; therefore, it is useful 
to express such combinations in terms of the 
$D^{*-} D_s^{*}$ branching fraction.

The contributions to the channels (\ref{channel1}),(\ref{channel2})
related to the radial excitations of negative and positive 
parity intermediate mesons can be estimated to represent
less than $15\%$ than the contribution of the corresponding low-lying states, 
using, e.g., the approach described in ref.\cite{Colangelo:1990rv}.

\begin{center}
\begin{figure}[htb]
\hspace*{0.5cm}
\psfig{figure=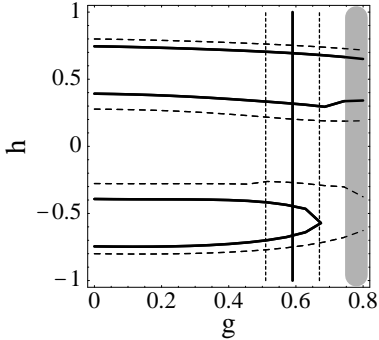,height=5.0 cm}
\vspace*{-0.4cm}
\caption{Allowed $1-\sigma$ (continuous lines) and $2-\sigma$ (dashed lines) 
regions in the $(g,h)$ plane. The vertical lines represent the result in 
ref.\cite{cleonew}, the shaded area corresponds to the bound 
in ref.\cite{accmor}.}
\label{fig:constraints}
\end{figure}
\end{center}

\begin{figure}[htb]
\hspace*{0.2cm}
\begin{tabular}{c c}
\psfig{figure=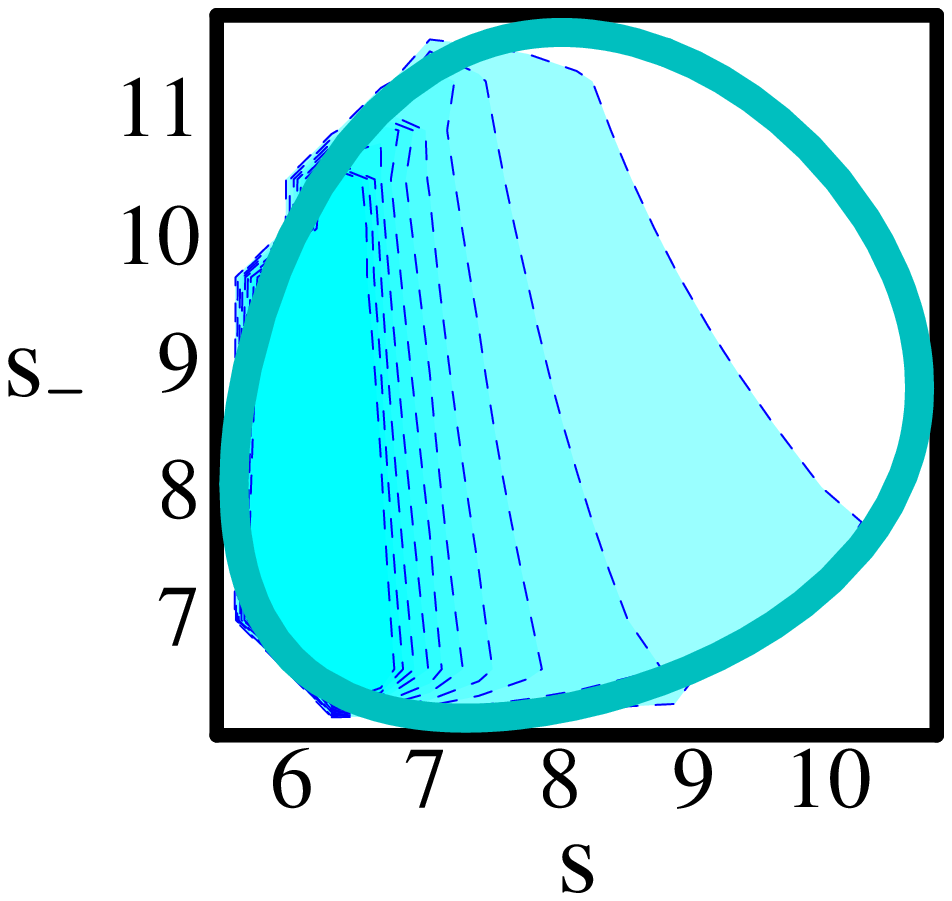,height=3.0 cm}&
\psfig{figure=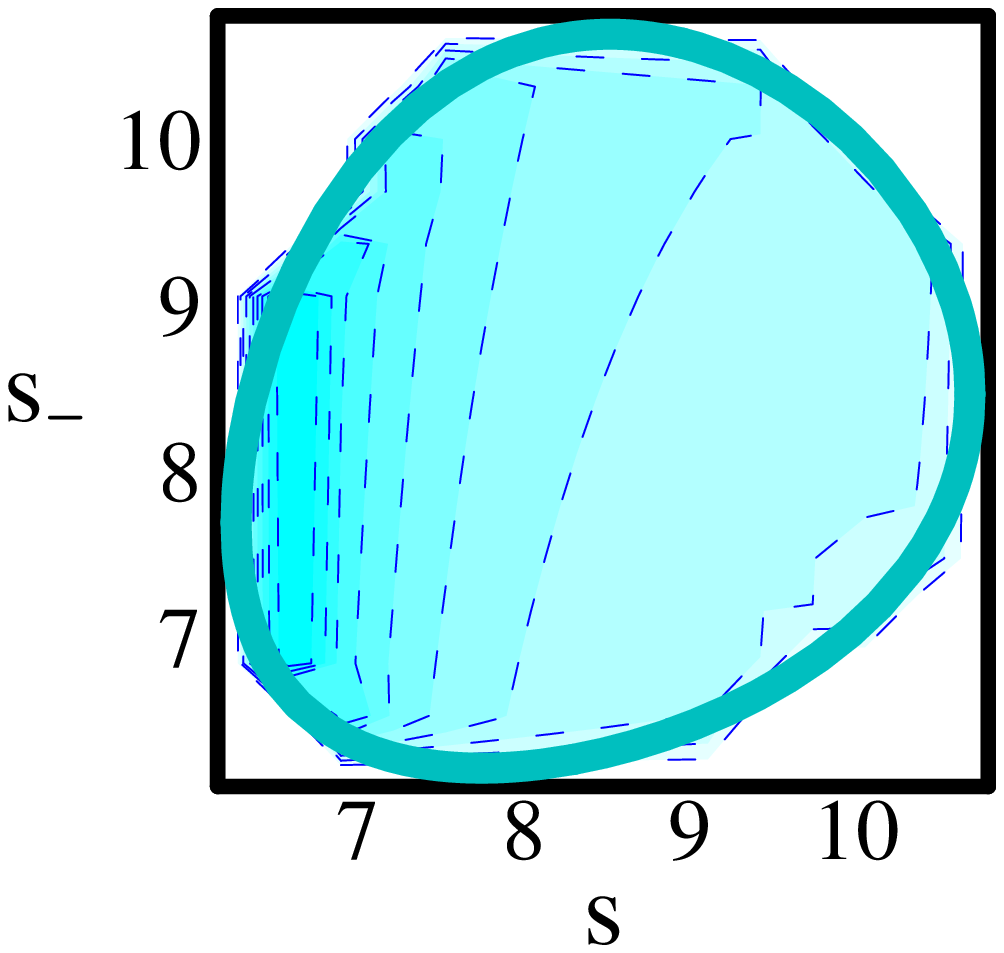,height=3.0 cm}\\
\psfig{figure=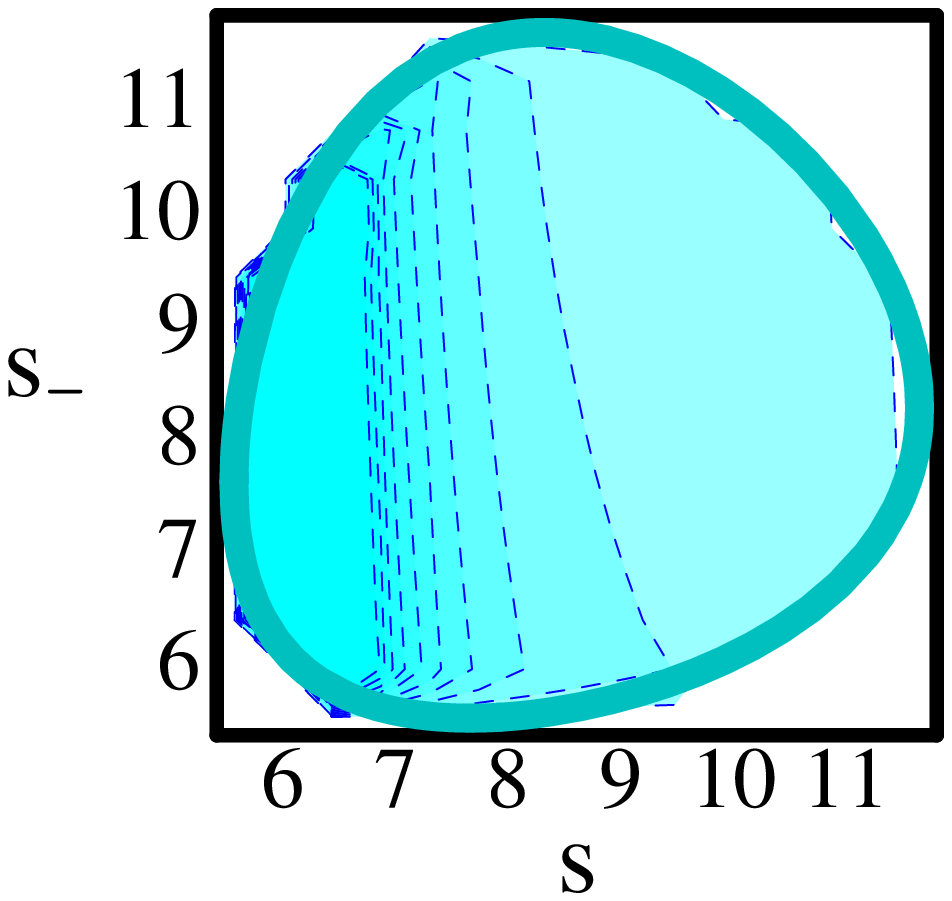,height=3.0 cm}&
\psfig{figure=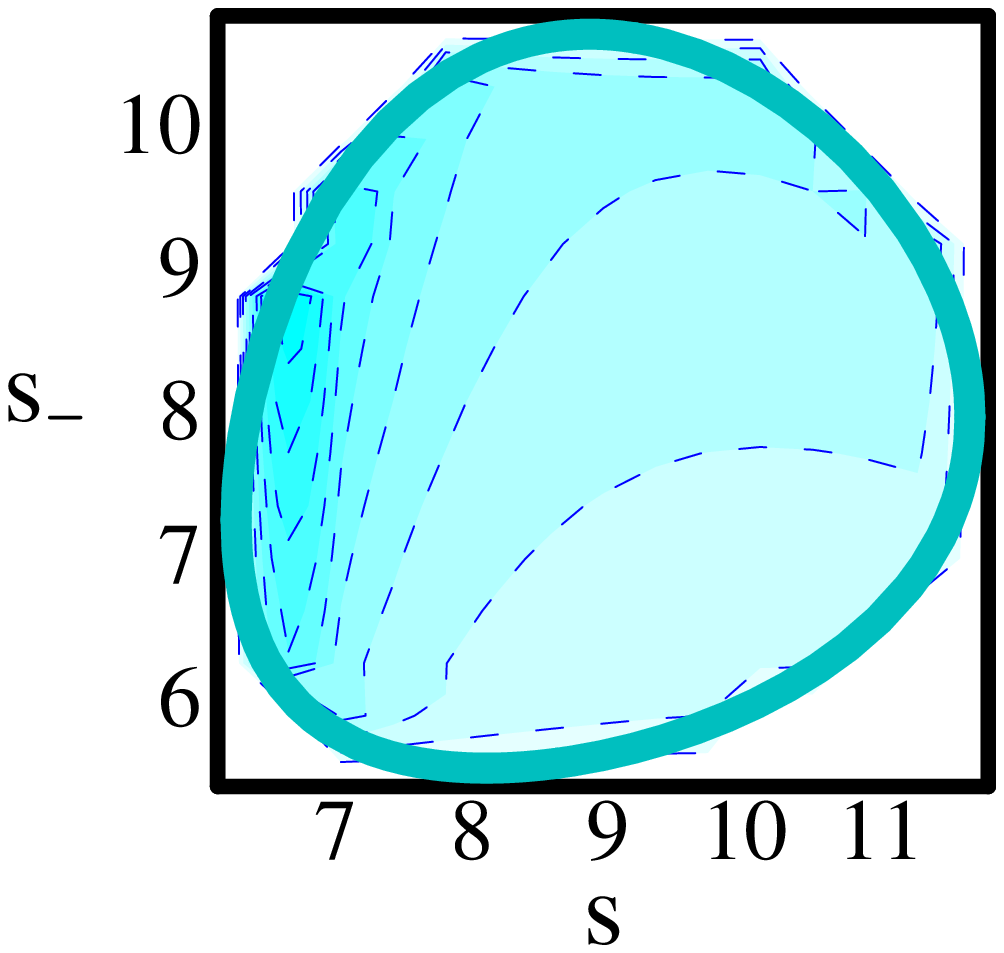,height=3.0 cm}\\
\end{tabular}
\vspace*{-0.4cm}
\caption{Dalitz plot of 
$B^0 \to D^{*-} D^0 K^+$ (up-left), $D^{*-} D^{*0} K^+$ (up-right),
$D^{-} D^0 K^+$ (down-left) and $D^{-} D^{*0} K^+$ (down-right). 
Units of $s=(p_{D^{(*)0}}+p_K)^2$ and  
$s_-=(p_{D^{(*)-}}+p_K)^2$ are GeV$^2$.}
\label{fig:DstarDK2}
\end{figure}

The measurement of the ratios
${\cal B}(B^0 \to D^{*-}  D^0  K^+)/{\cal B}(B^0 \to D^{*-}  D^*_s)$ and   
${\cal B}(B^0 \to D^{*-}  D^{*0}  K^+)/{\cal B}(B^0 \to D^{*-}  D^*_s)$ 
\cite{Aubert:2001wz,Groom:in} constrains 
the effective couplings  $g$ and $h$. The experimental central 
values are obtained for  $(g,h)=(0.05,-0.59)$ and  
$(g,h)=(0.0,+0.60)$. As depicted in fig.\ref{fig:constraints},
the allowed regions in the plane $(g,h)$
are tightly bounded along the $h$ direction, in the range
$|h|=0.6 \pm 0.2$ compatible with the expectations 
\cite{noidelta}, while the dependence on $g$ is mild, the range 
extending over all the presently permitted values
between $g=0$, the result $g=0.59 \pm 0.01 \pm 0.07$
obtained by the CLEO collaboration \cite{cleonew} and the upper bound $g<0.76$
from the ACCMOR collaboration \cite{accmor},\cite{khod}.
This means that the main contributions
to the processes in (\ref{channel1})  
are  not the $0^-$ and $1^-$, $D_s$ and $D^*_s$ intermediate states, 
but the positive parity $0^+$ and $1^+$ states $D_{s0}$ and 
$D_{s1}^*$, and therefore the  
$B^0 \to D^{*-} D^{0} K^+$ and $B^0 \to D^{*-} D^{*0} K^+$
decay modes are well suited for separately studying the properties of the 
$D_{s0}$  and $D_{s1}^*$ resonances by the analysis of Dalitz plots 
that are expected as depicted in fig.\ref{fig:DstarDK2}.
Broad resonances should be observed in the  $D^{(*)0} K^+$ invariant mass,
with  $\Gamma(D_{s0}) \simeq 180$ MeV and $\Gamma(D^*_{s1}) \simeq 165$ MeV.
The same structures are expected in the modes
$B^0 \to D^{-} D^{0} K^+$ and $B^0 \to D^{-} D^{*0} K^+$. 

An interesting problem concerns the reduction of the two main 
uncertainties of the approach, i.e. the use of the heavy quark limit both for
beauty and charm quarks, and the factorization ansatz employed in the 
nonleptonic matrix elements.  From the theoretical
point of view, a quantitative assessment of the
role of such approximations is a nontrivial task 
and one has to analyze hints from the measurements. 
The decay modes  (\ref{channel1},\ref{channel2})  
are different from the processes for which factorization has been proved 
in the infinite $b$ mass limit \cite{bbns}. Nevertheless, for
channels of the class we are considering, i.e. color allowed B transitions 
to pairs of charmed mesons, the factorization model produces results 
in agreement with the available data, within 
the current experimental uncertainties \cite{rosner}. The approach
proposed here can be considered as a further test of factorization 
for three-body decays. 
The other main uncertainty is the use of the infinite mass limit
for the charm quark. Finite charm quark mass effects can be
at the origin of a deviation observed in the (preliminary) measurement
of the branching fractions of (\ref{channel2}) 
with respect to the results obtained in the infinite limit. 

\section{$B^0 (\overline{B^0}) (t) \to D^+ D^- \pi^0$
AND THE WEAK PHASE $\beta$}

At the B factories, the analysis of the time-dependent Dalitz plot of the
three-body modes $B^0 ({\overline {B^0}}) \to D^- D^+ \pi^0$ 
would be useful for the 
investigation of CP violating effects in neutral $B$ decays and, 
in particular, to get new information on the  weak phase $\beta$
\cite{Charles:1998vf}.
This can be easily understood, since
the time-dependent decay probabilities of states identified at $t=0$ as 
$B^0$ and $\overline{B^0}$ respectively read:
\begin{eqnarray}
|A(B^0(t) \to D^+ D^- \pi^0|^2 \sim G_0(s_+,s_-)
\;\;\;\;\;\; \;\;\;\;\;\; \;\;\;\;\;\; \nonumber \\
+G_c(s_+,s_-) \cos(\Delta m t) - G_s(s_+,s_-) \sin(\Delta m t) \,\, ,\nn \\
|A(\overline {B^0}(t) \to D^+ D^- \pi^0|^2 \sim G_0(s_-,s_+) 
\;\;\;\;\;\; \;\;\;\;\;\; \;\;\;\;\;\; \nonumber \\
-G_c(s_-,s_+) \cos(\Delta m t) + G_s(s_-,s_+) \sin(\Delta m t)] \,\, , \nn
\end{eqnarray}
with 
\bea
G_s(s_+,s_-) = &-& 2 \sin(2 \beta) \, {\rm Re} \tilde G_s(s_+,s_-) \nn \\
&+&2 \cos(2 \beta) \, {\rm Im} \tilde G_s(s_+,s_-)  \,\,\, .
\label{Gs}
\eea
Therefore, through the analysis of 
$B^0 ({\overline {B^0}})(t) \to D^- D^+ \pi^0$
one envisages the possibility 
to access both $\sin(2 \beta)$ and $\cos(2 \beta)$, and therefore to
solve the discrete ambiguity $\beta \to {\pi \over 2}- \beta$
still present in current measurements from, e.g., 
CP asymmetries in $B \to J/\psi (\psi(2S), \dots) K_S$. 
This is an important issue, if one wants to
answer the question whether the weak angles measured in experiments that 
test CP violation match the angles that are being determined by measuring
quantities that conserve CP, e.g. the lenghts of the sides of the 
unitarity triangle \cite{parodi}. 

\begin{figure}[htb]
\hspace*{1.5cm}
\begin{picture}(80,2)(0,0)
\put(1,-30){$\displaystyle{{{\rm Re} \, \tilde G_s}}$}
\put(1,-130){$\displaystyle{{{\rm Im} \, \tilde G_s}}$}
\end{picture}
\begin{center}
\begin{tabular}{c}
\psfig{figure=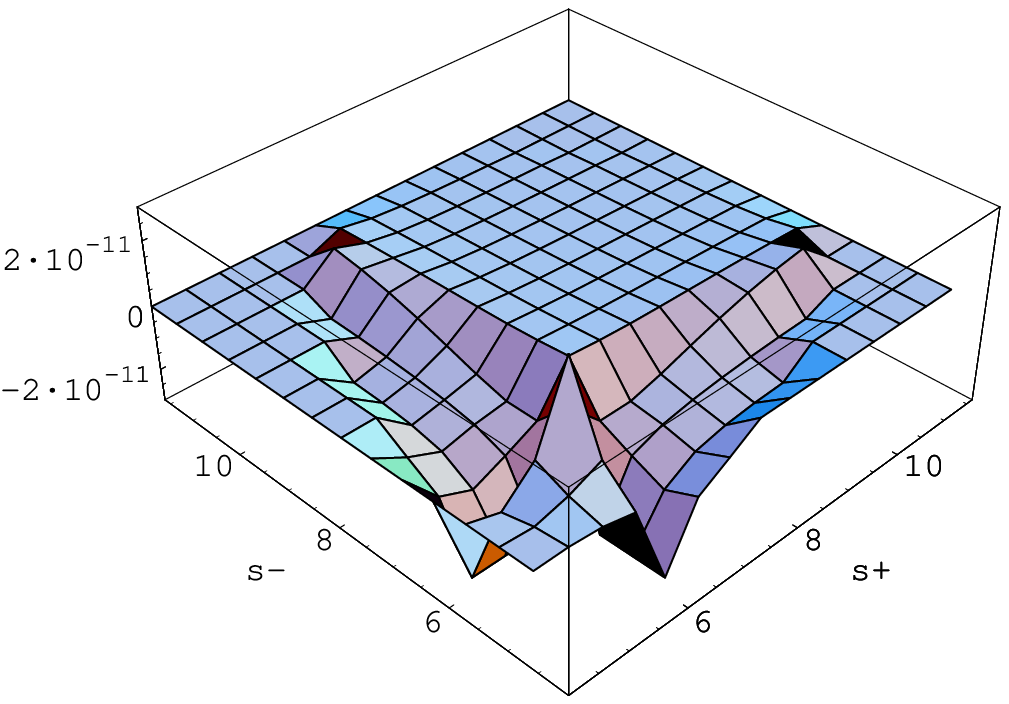,height=3.5 cm}\\
\vspace*{0.3cm}
\psfig{figure=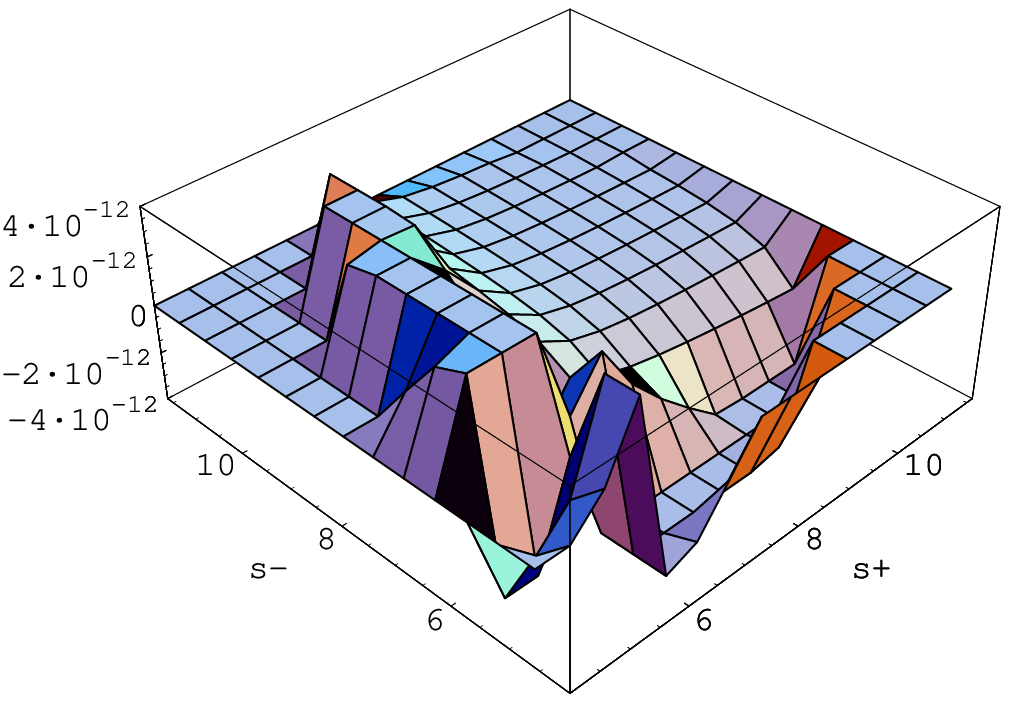,height=3.5 cm}\\
\end{tabular}
\end{center}
\vspace*{-0.4cm}
\caption{${\rm Re}\, \tilde G_s(s_+,s_-)$ (up) and
${\rm Im}\, \tilde G_s(s_+,s_-)$ (down) for the transition 
$B^0 \to D^+ D^- \pi^0$. Units of $s_\pm=(p_{D^\pm}+p_{\pi^0})^2$ are GeV$^2$.}
\label{fig:cp}
\end{figure}

The measurement is feasible:  
various amplitudes with different strong phases, corresponding to 
the decay chain 
$B^0 (\overline {B^0}) (t) \to D^+ D^{**-}+ D^- D^{**+}+
\dots \to D^+ D^- \pi^0$, contribute to the final state
producing sizeable interferences in the Dalitz plot. 
The coefficient of $\cos (2 \beta)$ in (\ref{Gs}),
${\rm Im} \tilde G_s$,  depicted in fig.\ref{fig:cp},
is expected to be different from zero
over a large portion of the Dalitz plot, thus allowing the identification of 
the $\cos (2\beta)$ term.
The application of the method to other three-body  modes
has been investigated in ref.\cite{Charles:1998vf}.

\noindent {\bf ACKNOWLEDGMENTS} \\
\noindent I thank F.~De Fazio, G.~Nardulli, N.~Paver and
Riazuddin for collaboration on the issues discussed in this talk.

\end{document}